\documentclass[11pt,a4paper,oneside]{article}

\begin{document}

\title{Some relations between hadron masses}

\author{T.\thinspace Jacobsen\\
Department of Physics, University of Oslo\\
PB 1048 Blindern\\
N-0316 Oslo, Norway}
\maketitle

\begin{abstract}
The mass of some hadrons are reproduced in terms of the mass of the nucleon.
A possible reason for emission of soft gammas is proposed.
\end{abstract}

\bigskip
\begin{flushleft}
\end{flushleft}
\thispagestyle{empty}

\newpage

\noindent
During the last 4-5 decades, many new mesons and baryons have been observed 
\cite{plb}. In this note we show that some of their masses probably are 
related, and that some phenomenological formulas seem to reproduce some hadron 
masses fairly well.

According to the quark model, mesons are quark-antiquark systems $q \bar{q}$. 
For each possible $q \bar{q}$ combination, the minimum mass meson \cite{plb} 
is listed in Table\thinspace I. In the following, masses are written in MeV/c$^2$.

\begin{center}{\large {\bf Table I}}\end{center}
\begin{center}
\begin{tabular}{|c|c|c|c|c|c|c|} \hline
&$\bar{u}$&$\bar{d}$&$\bar{s}$&$\bar{c}$&$\bar{b}$&$\bar{t}$ \\ \hline
$u$&$\pi^0 (135)$&$\pi^+ (140)$&$K^+ (494)$&$\bar{D}^0 (1865)$&$B^+ (5279)$& \\
$d$&$\pi^- (140)$&$\pi^0 (135)$&$K^0 (498)$&$D^- (1869)$&$B^0 (5279)$& \\
$s$&$K^- (494)$&$\bar{K}^0 (498)$&$\phi (1020)$&$D_s^- (1968)$&$B_s^0 (5370)$& \\
$c$&${D}^0 (1865)$&$D^+ (1869)$&$D_s^+ (1968)$&$\eta_c (2980)$&$B_c (6400)$& \\
$b$&$B^- (5279)$&$\bar{B}^0 (5279)$&$\bar{B}_s^0 (5370)$&$B_c (6400)$&$\Upsilon(9460)$&\\
$t$& & & & & & \\ \hline
\end{tabular}
\end{center}

\noindent
It is seen from Table\thinspace I that
\begin{eqnarray*}
m(\phi) + m(D_s) &=& 1020 + 1968 \;\;\approx\;\; m(\eta_c), \\
m(\phi) + m(B_s) &=& 1020 + 5370 \;\;\approx\;\; m(B_c), \\
m(\eta_c) + m(B_c) &=& 2979 + 6400 \;\;\approx\;\; m(\Upsilon),
\end{eqnarray*}
which suggest that the masses of some mesons are correlated.\\

The relation
\begin{center}
$m(N) / 2 \pi \;\; \approx\;\; 150\;$MeV/c$^2\;\;\approx\;\; m(\pi)$
\end{center}
\noindent
\cite{jt}, which reproduces the mass $m(\pi)$ of the pion in terms of the mass $m(N)$ 
of the nucleon fairly well, corresponds to
\begin{center}
$\lambda_c (\pi) \;\; \approx\;\; 2 \pi \lambda_c (N)$
\end{center}
\noindent
where $\lambda_c(\pi)$ and $\lambda_c(N)$ are the the Compton wave lengths of the pion 
and the nucleon, respectively, which suggests a geometric relation between the two 
Compton wave lenghts.\\

By means of an extra factor $(4\pi)^{n/2}$ on the left hand side of the mass formula 
above, \cite{jt}, the trend of increase of mass along the quasi-diagonal
\begin{center}
$u \bar{d}, d \bar{s}, s \bar{c}, c \bar{b}$
\end{center}
\noindent
in Table\thinspace I is fairly well reproduced for $n = 0,1,2,3,$ i.e.
\begin{eqnarray*}
(4\pi)^{0/2}\; m(N)/2\pi &\approx&\;\: 150 \;\;\approx\;\; m(\pi), \\
(4\pi)^{1/2}\; m(N)/2\pi &\approx&\;\: 530 \;\;\approx\;\; m(K), \\
(4\pi)^{2/2}\; m(N)/2\pi &\approx& 1880 \;\;\approx\;\; m(D), \\
(4\pi)^{3/2}\; m(N)/2\pi &\approx& 6660 \;\;\approx\;\; m(B_c).
\end{eqnarray*}

\noindent
If also
\begin{center}
$(4\pi)^{4/2}\; m(N)/2\pi \;\; \approx\;\; 23610 \;\;\approx\;\; m(b \bar{t})$
\end{center}
\noindent
and by means of an extrapolation of the sums above, one might expect
\begin{center}
$m(\Upsilon) + m(b\bar{t}) \;\; \approx\;\; 9460 + 23610\;\; 
\approx\;\; 33070 \;\;\approx\;\; m(t \bar{t})$.
\end{center}

\noindent
By means of the ratio
\begin{center}
$r \;\;=\;\; m/m(\pi)$
\end{center}
\noindent
we measure the mass of some mesons in $m(\pi)$-units. $r$ and the value of  
 \begin{center}
$2^k \pi^{n/2}$
\end{center}
\noindent
which for integer $k$ and $n$ gives the best fit to $r$ are listed in Table\thinspace II.

\begin{center}{\large {\bf Table II}}\end{center}
\begin{center}
\begin{tabular}{|c|c|c|c|l|} \hline
Meson& $I$ & $J^P$ & $r = m/m(\pi)$ & $2^k \pi^{n/2}$ \\ \hline
$\eta$ & 0   & $0^-$ & 547/140 = 3.91 & $2^2 \pi^{0/2} = 4.00$ \\
$K$    & 1/2 & $0^-$ & 497/140 = 3.55 & $2^1 \pi^{1/2} = 3.54$ \\
$\omega$ & 0 & $1^-$ & 782/140 = 5.59 & $2^0 \pi^{3/2} = 5.56$ \\
$K^*$  & 1/2 & $1^-$ & 892/140 = 6.37 & $2^1 \pi^{2/2} = 6.28$ \\
$J/\psi$ & 0 & $1^-$ & 3097/140 = 22.12 & $2^2 \pi^{3/2} = 22.26$ \\ \hline
\end{tabular}
\end{center}
\noindent
Acording to this table, these mesonic masses are fairly well reproduced by the formula 
\begin{center}
$m \;\;\approx\;\; 2^k \pi^{n/2}\; m(\pi)$
\end{center}
\noindent
in favour of some mutual relationship between these mesons.

While the mass differences between the non-strange, strange, double and triple strange 
baryons in the well known baryon decouplet is about 140 MeV/c$^2$, which is the mass of 
the pion, the baryons in the octet do not show the same differences. While the difference 
between the mass of the $\Delta$(1232) and the mass of the nucleon is 290 MeV/c$^2$, the 
difference between the mass of the $\Lambda$(1115) and the mass of the nucleon is 
175 MeV/c$^2$. \\

If every baryon heavier than the nucleon is taken to be an excited state of 
a nucleon, then the relative excitation energies would be important. 
In Tables\thinspace III-IV we therefore compare the successive mass differences $d$ with 
the smallest one for baryons with $J^P = 1/2^+$, and $J^P = 3/2^+$, \cite{plb}, 
respectively.

\begin{center}{\large {\bf Table III}}\end{center}
\begin{center}
\begin{tabular}{|c|c|c|c|c|c|} \hline
Baryon  & $I$ & $J^P$ & $m$ & $d \simeq m-m(N)$&$d/175\simeq 2^{n/k}$ \\ \hline
$\Lambda$ & 0   & $1/2^+$ &1116 & 175 & $1.0 = 2^{0/2}$ \\
$\Sigma$ & 1   & $1/2^+$ &1197 & 255 & $1.4 \approx 2^{1/2}$ \\
$\Xi$ & 1/2   & $1/2^+$ &1315 & 375 & $2.1 \approx 2^{2/2}$ \\
$N$ & 1/2   & $1/2^+$ &1440 & 500 & $2.9 \approx 2^{3/2}$ \\
$N$ & 1/2   & $1/2^+$ &1640 & 700 & $4.0 = 2^{4/2}$ \\ \hline
\end{tabular}
\end{center}
\noindent
For these five $J^P = 1/2^+$ baryons, the mass difference $d$ is for $k = 2$ and 
\mbox{$n$ = 0, 1, 2, 3, or 4}, fairly well described by
\begin{center}
$d \;\;\approx\;\; 2^{n/k}\; 175\;$MeV/c$^2$.
\end{center}

\begin{center}{\large {\bf Table IV}}\end{center}
\begin{center}
\begin{tabular}{|c|c|c|c|c|c|} \hline
Baryon  & $I$ & $J^P$ & $m$ & $d \simeq m-m(N)$&$d/290\simeq 2^{n/k}$ \\ \hline
$\Delta$ & 3/2 & $3/2^+$ &1232 & 290 & $1.0 = 2^{0/3}$ \\
$\Sigma^*$ & 1 & $3/2^+$ &1385 & 445 & $1.5 \approx 2^{2/3}$ \\
$\Xi^*$ & 1/2 & $3/2^+$ &1530 & 590 & $2.0 \approx 2^{3/3}$ \\
$\Omega^-$ & 0 & $3/2^+$ &1672 & 730 & $2.5 \approx 2^{4/3}$ \\ \hline
\end{tabular}
\end{center}
\noindent
For these four $J^P = 3/2^+$ baryons, the mass difference $d$ is for $k = 3$ and 
\mbox{$n$ = 0, 2, 3, or 4}, fairly well described by 
\begin{center}
$d \;\;\approx\;\; 2^{n/k}\; 290\;$MeV/c$^2$.
\end{center}
\noindent

A $J^P = 3/2^+$  baryon with mass about 1308 MeV/c$^2$ corresponding to $2^{1/3}$ seems 
however to be missing according to this table. If such a baryon exists, it is possibly 
overlooked in the background below the $J^P = 3/2^+$ $\Xi(1315)$ peak.\\

\noindent
Table\thinspace V  shows the ratio
\begin{center}
$R\;\; =\;\; m/m(N)$
\end{center}
\noindent
for some baryons with mass $m$ and $F$-values defines by the formula 
\begin{center}
$F \;\;=\;\; (1 + 1/n)^2$
\end{center}
\noindent
for integer $0 < n < 12$.

\newpage
\begin{center}{\large {\bf Table V}}\end{center}
\begin{center}
\begin{tabular}{|c|c|c|r|r|} \hline
Baryon & $I$ & $J^P$ & $R=m/m(N)$ & $F=(1+1/n)^2$ \\ \hline
$\Lambda$ & 0 & $1/2^+$ &1116/940 = 1.19 & $(12/11)^2=1.19$ \\
\multicolumn{4}{|c|}{no baryon seen at 1137 corresponding to} & $(11/10)^2=1.21$ \\
\multicolumn{4}{|c|}{no baryon seen at 1160 corresponding to} & $(10/9)^2=1.23$ \\
$\Sigma$ & 1 & $1/2^+$ &1197/940 = 1.27 & $(9/8)^2=1.27$ \\
$\Delta$ & 3/2 & $3/2^+$ &1232/940 = 1.31 & $(8/7)^2=1.31$ \\
$\Xi$ & 1/2 & $1/2^+$ &1315/940 = 1.40 & $(7/6)^2=1.36$ \\
$\Sigma^*$ & 1 & $3/2^+$ &1385/940 = 1.47 & $(6/5)^2=1.44$ \\
N & 1/2 & $1/2^+$ &1440/940 = 1.53 & $(5/4)^2=1.56$ \\
$\Omega^-$ & 0 & $3/2^+$ &1672/940 = 1.78 & $(4/3)^2=1.78$ \\
$\Lambda$ & 0 & $5/2^+$ &2110/940 = 2.24 & $(3/2)^2=2.25$ \\
\multicolumn{2}{|c|}{uncertain} & &3800/940 = 4.04 & $(2/1)^2=4.00$ \\ \hline
\end{tabular}
\end{center}
\noindent

\noindent
In the region between $\Lambda$(1115) and $\Sigma$(1190), two non-observed
baryons are expected to exist according to this formula, as seen from 
Table\thinspace V. If their widths are large enough, they might have been taken to 
belong to the background in this region. While the geometric constant $\pi$ is 
needed for the reproduction of mesonic masses, $\pi$ is not needed for reproduction 
of the baryonic masses. The formula 
\begin{center}
$m \;\;\approx\;\; F \; m(N)$,
\end{center}
\noindent
thus reproduces some baryonic masses as if these baryons belong to a common chain.
It is interesting to note that for
\begin{center}
$F \rightarrow 1$,
\end{center}
\noindent
i.e. for decreasing mass $m$, the density of states increases and tends to appear 
as a continuum, contrary to e.g. hydrogen atoms where continuum appears for high 
excitations. For large $n$-values corresponding to a $Q$-value
\begin{center}
$Q \;\;<\;\; m(\pi)c^2$,
\end{center}
\noindent
i.e. too small for emissions of a pion, only $\gamma$ emission would be 
energetically possible. Since final state non-Bremsstrahlung soft $\gamma$'s 
have been observed in high energy $pp$ reactions \cite{french2}, such emission 
is a candidate for $\gamma$ decay of large $n$ baryons. Emission of 
non-Bremsstrahlung soft $\gamma$'s has also been observed in high energy 
$K^+ p$-reactions \cite{chli,bott}, $\pi^+ p$-reactions \cite{bott},
and in $\pi^- p$-reactions \cite{bane,belog,french1}. This suggests that not 
only nucleons but also mesons may be excited to energy levels too low for 
mesonic decays.

In summary, based only on the mass of the nucleon, some relations are found 
which reproduce some hadronic mass values fairly well. A possible reason for 
emission of soft gammas in hadron-proton collisions is proposed. 

\bigskip
\begin{center}{\Large {\bf Acknowledgement}}\end{center}
\noindent
Discussions with K.M.\thinspace Danielsen, University of Oslo, Norway, is 
greatefully acknowledged.

\end{document}